\documentclass{article}
\usepackage{spconf,amsmath,graphicx}

\usepackage{enumitem}
\setlist{nosep, leftmargin=14pt}
\usepackage[left=1.62cm,right=1.62cm,top=1.9cm]{geometry}
\usepackage{amsfonts} 
\usepackage{multirow}
\usepackage{pifont}
\usepackage{textcomp}

\title{Towards Zero-Shot Task-Generalizable Learning on fMRI}
\name{Jiyao Wang$^{1}$  Nicha C. Dvornek$^{1, 2}$  Peiyu Duan$^{1}$  Lawrence H. Staib$^{1, 2, 3}$  James S. Duncan$^{1, 2, 3}$}

\address{ 
$^{1}$ 
Department of Biomedical Engineering, Yale University, USA\\
$^{2}$ 
Radiology \& Biomedical Imaging, Yale School of Medicine, USA \\ 
$^{3}$ 
Electrical Engineering, Yale University, USA}
\begin{document}
\maketitle
\begin{abstract}
Functional MRI measuring BOLD signal is an increasingly important imaging modality in studying brain functions and neurological disorders. It can be acquired in either a resting-state or a task-based paradigm. Compared to resting-state fMRI, task-based fMRI is acquired while the subject is performing a specific task designed to enhance study-related brain activities. Consequently, it generally has more informative task-dependent signals. However, due to the variety of task designs, it is much more difficult than in resting state to aggregate task-based fMRI acquired in different tasks to train a generalizable model. To resolve this complication, we propose a supervised task-aware network TA-GAT that jointly learns a general-purpose encoder and task-specific contextual information. The encoder-generated embedding and the learned contextual information are then combined as input to multiple modules for performing downstream tasks. We believe that the proposed task-aware architecture can plug-and-play in any neural network architecture to incorporate the prior knowledge of fMRI tasks into capturing functional brain patterns.   

\end{abstract}
\section{Introduction}
\let\thefootnote\relax\footnotetext{\textcopyright 2025 IEEE.  Personal use of this material is permitted.  Permission from IEEE must be obtained for all other uses, in any current or future media, including reprinting/republishing this material for advertising or promotional purposes, creating new collective works, for resale or redistribution to servers or lists, or reuse of any copyrighted component of this work in other works.}

Recently, the rapid development of open access datasets has significantly alleviated the data insufficiency issue in medical imaging analysis. However, it also poses challenges in efficiently analyzing information in large amounts of data. Within the functional magnetic resonance imaging (fMRI) domain, open access datasets such as the Human Connectome Project~(HCP)~\cite{hcp} provide researchers with access to large-scale standardized fMRI data in both resting-state and task-based acquisition. 

Although constructing a data set from the resting state or any single task is straightforward, the resting-state assumption or any task-specific context may cause a strong inductive bias in training a neural network, impairing the model's capability to generalize on different tasks for a general understanding of brain functions. Moreover, from the perspective of the scaling law~\cite{scaling-law}, the performance of a neural network is generally positively correlated with the number of data samples. Using only a small portion of the entire dataset is a waste of information, especially in studies that attempt to develop a general-purpose foundation model. 

In line with this observation, Caro et al.~\cite{brainLM} have developed an fMRI foundation model using all resting-state and task-based data without distinguishing contextual differences during fMRI acquisition. Although their experiments show impressive performance in predicting masked fMRI segments, we argue that direct aggregation may not be the best approach to utilize fMRI data including various task acquisition paradigms. Incorporating task-awareness into network architecture may improve its performance in known heterogeneous tasks and zero-shot generalization capability in an unseen task.

Therefore, we propose the Task-Aware Graph Attention Network~(TA-GAT) to demonstrate the benefit of a task-aware learning strategy in task-based fMRI data. In summary, our main contributions are as follows:
\begin{itemize}
    \item We propose a novel task-aware graph neural network (GNN) framework to improve the model generalization on both seen and unseen task-based fMRI data. We believe it is also applicable as an add-on to most existing neural network architectures. 
    \item Based on the proposed framework, we show a novel application of orthogonal projection loss~\cite{ortho} to regularize the learning of task representations, which further improves the robustness of the network.
\end{itemize}

\begin{figure*}[t]
\centering
\includegraphics[width=0.9\textwidth]{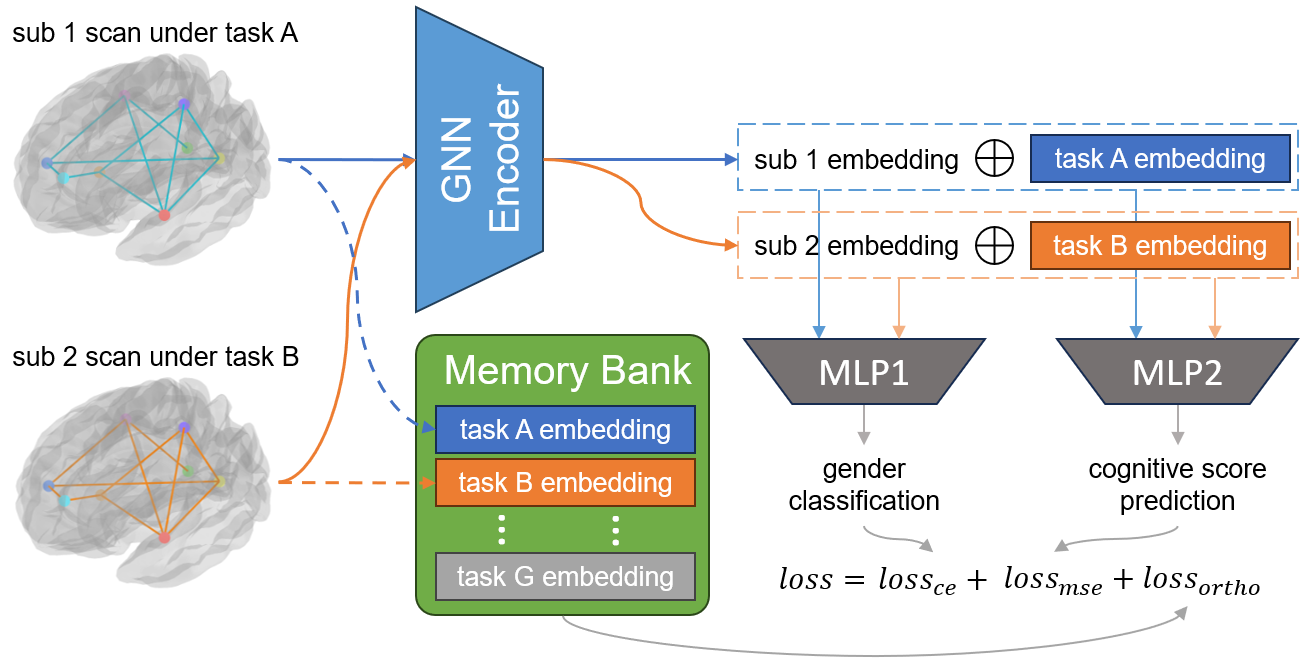}
\caption{Model architecture of TA-GAT. Blue and orange arrows denote the processing of two fMRI from distinct subjects and task stimuli. Task A to G corresponds to the 7 different fMRI tasks in HCP dataset.} \label{fig_architect}
\end{figure*}

\section{Data}
We use the preprocessed 3T task-based fMRI data in WU-Minn HCP 1200 Subjects Data Release~\cite{hcp} containing 7 different tasks: emotion, gambling, language, motor, relational, social, and working-memory. We filter out subjects with missing task scans or cognitive measurements, resulting in a total of 1008 subjects. For supervised multitask training, we use gender~(as labeled in HCP data dictionary) and age-adjusted total cognitive function composite scores as labels, with the cognitive scores normalized into range $[0,1]$. In cross-validation experiments, we divide the 1008 subjects into 5 roughly equal partitions.  
\section{Method}
\subsection{Notation and Problem Definition}
For each task scan of each subject, we construct a functional brain graph $G=(V,E)$ on the vertex set $V$ and the edge set $E$. A vertex set $V$ of $N$ graph nodes has node features $\{x_j\in\mathbb{R}^{d_{in}} | j\in[1,2,\ldots,N]\}$, where $d_{in}$ is the dimension of input node features. For any edge in the edge set $E$ that connects node $i$ and node $j$, we define the edge weight $w\in\mathbb{R}$ and $w>0$. For $M$ distinct fMRI tasks, we define a learnable representation vector $h_k\in \mathbb{R}^{d_{mem}}$, with $k\in\{1,2,\ldots,M\}$. The set of all task representations is our memory bank. Following the above definition, the multitask objective of performing the gender classification and normalized cognitive score prediction can be described as:
$$ f:(G,h_k)\mapsto (Y\in\{0,1\}, Z\in[0,1])$$
\subsection{Graph Construction}
For each fMRI scan, we parcellate the brain image into 268 regions of interest (ROIs) following the Shen atlas~\cite{shen268}. For the construction of nodes and edges, we adopt a strategy similar to the method described in Li et al.~\cite{BrainGNN}. Using the averaged fMRI signals in each ROI over all voxels, we compute Pearson's correlations of each node's signal with all the other nodes and use them as node features. Meanwhile, the top 5\% of the partial correlations is used to define the weighted edges.  

\subsection{Network Architecture}
As shown in Fig.~\ref{fig_architect}, there are 3 components in the proposed TA-GAT architecture: 1) a GNN encoder, 2) a memory bank, and 3) two MLPs, one for each task.
\subsubsection{General-purpose GNN Encoder}
The proposed general-purpose GNN encoder is composed of two consecutive GAT~\cite{gat} convolution layers. On the updated output node features in each layer, we apply both global mean-pooling and global max-pooling operations to extract permutation-invariant features. The outputs from the four pooling operations (two in each layer) are concatenated as the output embedding. 
\subsubsection{fMRI Task Memory Bank}
Keeping a memory bank of parameter representations depicting each instance is a frequently applied architecture in representation learning~\cite{mem-bank}. In our application, instead of keeping a parameter representation for each fMRI scan instance, we only keep seven learnable categorical representations for all distinct fMRI tasks. For each graph input, after the GNN encoder generates the general-purpose embedding, we pass the corresponding task representation to a single-layered projection and concatenate the GNN-generated embedding with the output from the projection layer. 
\subsubsection{Downstream MLP}
In the proposed design, we utilize the concatenated instance-level GNN-generated information with the group-level task representation stored in the memory bank for two downstream tasks: 1) classification of gender, 2) prediction of total cognitive function composite scores. For each aim, we train an MLP module to extract different information from the same general-purpose embedding. Using these two metrics as examples, we aim to show that the generated embedding contains both demographic and psychometric information.

\subsection{Loss Functions}
We use a weighted sum of 3 different loss terms in training the network. For gender classification, we apply the cross-entropy loss $L_{ce}$~(Eq.~\ref{loss_ce}). For the prediction of total cognitive function composite scores, we measure a MSE loss $L_{mse}$~(Eq.~\ref{loss_mse}). To regularize the task representations in the memory bank, we apply an orthogonal projection loss $L_{ortho}$~(Eq.~\ref{loss_ortho}), which computes the cosine similarity between the different task representations and encourages them to be orthogonal. Similar constructions of this loss have previously been proposed in the image classification task to regularize the representation of images~\cite{ortho} and in attention algorithm to regularize attention weights~\cite{ortho-demo}. 
\begin{equation}
\mathcal{L}_{ce} = -\frac{1}{N} \sum_{i=1}^{N} \left[ y_i \log(\hat{y}_i) + (1 - y_i) \log(1 - \hat{y}_i) \right]
\label{loss_ce}
\end{equation}

\begin{equation}
\mathcal{L}_{mse} = \frac{1}{N} \sum_{i=1}^{N} (z_i - \hat{z}_i)^2
\label{loss_mse}
\end{equation}

\begin{equation}
\mathcal{L}_{ortho} = \frac{1}{M(M-1)} \sum_{p=1}^{M} \sum_{\substack{q=1, \\ q \neq p}}^{M} \frac{h_p \cdot h_q}{\|h_p\|_2 \|h_q\|_2}
\label{loss_ortho}
\end{equation}

\begin{equation}
\mathcal{L} = \mathcal{L}_{ce}+\lambda_1\cdot\mathcal{L}_{mse}+\lambda_2\cdot\mathcal{L}_{ortho}
\end{equation}

\begin{table}[t]
\centering
\setlength{\tabcolsep}{3mm}
\begin{tabular}{c|c c c c c c c}
{} & \textbf{A} & \textbf{B} & \textbf{C} & \textbf{D} & \textbf{E} & \textbf{F} & \textbf{G}\\
\hline
\textbf{1} & \ding{51} & \ding{51}& \ding{51}& \ding{51}& \ding{51}& \ding{51}& {}\\
\textbf{2} & \ding{51} & \ding{51}& \ding{51}& \ding{51}& \ding{51}& \ding{51}& {}\\
\textbf{3} & \ding{51} & \ding{51}& \ding{51}& \ding{51}& \ding{51}& \ding{51}& {}\\
\textbf{4} & \ding{51} & \ding{51}& \ding{51}& \ding{51}& \ding{51}& \ding{51}& {}\\
\textbf{5} & \ding{55} & \ding{55}& \ding{55}& \ding{55}& \ding{55}& \ding{55}& \ding{108}\\
\end{tabular}
\caption{Illustration of experiment setup. Rows 1-5 denote cross-validation data partitions. Columns A-G denote distinct fMRI tasks. \ding{51} denote data used as training. \ding{55} and \ding{108} denote data used as known-task testing and unseen-task testing respectively.}
\label{tab_setup}
\end{table}

\section{Experiments and Results}

\begin{table*}[t]
\centering
\setlength{\tabcolsep}{0.5mm}
\begin{tabular}{c | c | c | c c c c c c c }
\hline
\multicolumn{3}{c}{} & Emotion & Gambling & Language & Motor & Relational & Social & WM\\ 
\hline
\hline
{} & \multirow{2}*{w/o TA} & Acc(\%) & 78.8(1.9) & 79.1(2.2) & 80.3(2.1) & 78.7(2.2) & 79.3(1.6) & 79.0(0.5) & 81.4(1.5)\\
{Known} & & Corr & 0.237(0.033) & 0.244(0.028) & 0.254(0.036) & 0.236(0.027) & 0.232(0.031) & 0.226(0.021) & 0.210(0.024)\\
\cline{2-10}
{tasks}& \multirow{2}*{TA-GAT} & Acc(\%) & \textbf{82.4(0.9)} & \textbf{81.9(1.1)} & \textbf{82.0(1.2)} & \textbf{82.0(1.8)} & \textbf{81.4(1.0)} & \textbf{82.9(0.5)} & \textbf{83.0(1.3)}\\
{}& & Corr & \textbf{0.282(0.035)} & \textbf{0.256(0.026)} & \textbf{0.262(0.051)} & \textbf{0.269(0.027)} & \textbf{0.258(0.026)} & \textbf{0.245(0.027)} & \textbf{0.256(0.032)}\\
\hline
{}& \multirow{2}*{w/o TA} & Acc(\%) & 78.1(3.2) & 79.2(3.3) & 79.1(3.4) & 76.5(3.5) & 79.2(2.2) & 78.2(4.8) & \textbf{77.9(2.4)}\\
{Unseen}& & Corr & 0.165(0.041) & 0.211(0.053) & 0.220(0.029) & 0.131(0.093) & 0.246(0.048) & 0.232(0.044) & 0.146(0.034)\\
\cline{2-10}
{task}& \multirow{2}*{TA-GAT} & Acc(\%) & \textbf{82.2(2.9)} & \textbf{82.1(2.6)} & \textbf{79.9(3.1)} & \textbf{78.9(1.4)} & \textbf{81.3(1.8)} & \textbf{80.3(1.7)} & 76.5(2.9)\\
{}& & Corr & \textbf{0.206(0.070)} & \textbf{0.235(0.056)} & \textbf{0.228(0.053)} & \textbf{0.160(0.071)} & \textbf{0.265(0.045)} & \textbf{0.273(0.065)} & \textbf{0.156(0.021})\\
\hline

\end{tabular}
\caption{Classification accuracy~(acc) of gender and Pearson's correlation~(corr) between cognitive score prediction and ground truth on test-subject data under known tasks and the unseen task. We compare the performance of TA-GAT and the same architecture without the task-aware memory bank. The numbers are reported in mean(std) with the best performance in known-task test and unseen-task test bolded.}
\label{tab_exp}
\end{table*}
\subsection{Experimental Setup}
In experiments, we set 2048 as the width of the hidden dimension and the output dimension of the GNN layers, leading to an 8192-dimensional embedding after pooling operations. For the memory bank, we also choose 2048 as the embedding dimension for contextual information of each distinct fMRI task. The two MLPs have an identical structure except for the output layer. For classification, the two-channel output is passed to a softmax operation. For prediction, sigmoid activation is used to limit the single-channel output to between 0 and 1. Each MLP has 4 layers with SiLU~\cite{silu} activation and a dropout rate of 0.2 in each layer. 

For the scalar weights of the loss terms, we choose $\lambda_1=50$ and $\lambda_2=1$. During training, we use an SGD optimizer along with a step learning rate scheduler. The initial learning rate is $4\times10^{-6}$ with a step size of 10 epochs and $\gamma=0.4$ on 100 epochs. The experiments are performed on a single NVIDIA A100 GPU.

\subsection{Main Results}

To evaluate the proposed architecture, we perform cross-validation experiments on the HCP task-based fMRI data and compare the performance in TA-GAT with or without the proposed task-aware memory bank. As illustrated in Table~\ref{tab_setup}, in each experiment, we train the network on 6 fMRI tasks and leave 1 fMRI task out to test the robustness of the model on an unseen task. The leave-one-out experiment is iterated through all 7 combinations of tasks. The results are listed in Table~\ref{tab_exp}.

The results show that the proposed task-aware architecture improves the network performance in both classification and prediction for all experiments except for one classification instance tested on the unseen working memory task. 

For the known tasks included in the training data, the knowledge of task categorization and the capability of performing task-dependent computation help make the model more robust on test subjects. The orthogonal projection loss also further regularizes the representation by encouraging the memory bank to store only task-specific context orthogonal to each other. 

Meanwhile, for the unseen task not included in the training data, although its representation is never updated by task-specific input data, it is regularized by the orthogonal projection loss and shows a reasonable zero-shot generalization performance. 

\begin{table}[t]
\centering
\begin{tabular}{ c | c | c c}
\multicolumn{2}{c|}{} & Acc(\%) & Corr \\
\hline
\hline
 {Known} &TA-GAT & 82.4(0.9) & 0.282(0.035) \\
 {tasks} & w/o $\mathcal{L}_{ortho}$  & 81.1(1.3) & 0.281(0.032) \\
 \hline
 {Unseen} &TA-GAT & 82.2(2.9) & 0.206(0.070) \\
 {task} & w/o $\mathcal{L}_{ortho}$  & 77.0(2.3) & 0.199(0.070) \\
 \hline
\end{tabular}
\caption{Ablation study on orthogonal projection loss using emotion task as unseen. Numbers are listed in mean(std).}
\label{tab_ablation}
\end{table}

\subsection{Ablation Study}
We perform an ablation study on the orthogonal projection loss. Using the emotion task as unseen, we run cross-validation experiments and compare the performance of the TA-GAT framework with the same architecture trained without $\mathcal{L}_{ortho}$. 

The results in Table~\ref{tab_ablation} show that although the two networks have similar performance in testing using known-tasks, the model trained without $\mathcal{L}_{ortho}$ has a worse performance when encountering unseen tasks, especially in the classification task. The observations support our claim that the orthogonal projection loss can help regularize the learnable representations of the task contexts. It not only improves network performance on known tasks from test subjects (81.1 to 82.4), but also significantly raises zero-shot generalization robustness to an unseen task (77.0 to 82.2). 

\section{Discussion}
\subsection{Why task-based fMRI?}
Resting-state fMRI is usually more preferrable in large-scale fMRI analysis due to its availability and homogeneous non-stimuli assumption. However, we would like to provide two reasons for choosing task-based fMRI.

fMRI is known as a noisy modality. In addition to the noise originating from data acquisition, the randomness of brain activities also contributes to the chaos. Supported by a growing amount of work in reconstructing viewed images from fMRI~\cite{fMRI-img}, there are likely at least two sides of brain activity that are reflected in fMRI: 1) the pattern of thinking and 2) the information contained in thoughts, for example, visual or vocal input from the environment. Under this assumption, distinguishing patterns from information can be helpful in extracting individual features by eliminating interference from irrelevant information. For task-based fMRI, we know exactly what the environmental inputs are. 

Practically, the GNN network proposed in this paper, as well as most of the existing GNN networks, relies on an accurate definition of graph edges to achieve optimal performance. As demonstrated in \cite{stnagnn}, the graph edges defined based on task-based fMRI are usually more stable.

\subsection{Why ROI-based graph?}
Recent advances in large pre-trained networks have also raised interest in the field of medical imaging to either adopt pre-trained networks into medical applications~\cite{sam-adopt} or develop medical foundation models~\cite{brainLM,resvit}. However, despite the fact that analyzing fMRI data using an ROI-based brain graph is a widely accepted method \cite{BrainGNN, zhao-stgnn}, there is still no foundation model trying to explain or simulate the connectivity pattern between brain ROIs. 

Compared to other approaches, the ROI-based graph has its unique advantage in explaining regional brain connectivity and detecting sub-networks, which are important for understanding brain functions.

Although currently in a supervised dual-task schema, we show the possibility of training a general-purpose encoder using ROI-based graphs as input. In future work, we will try to adopt similar designs to unsupervised pre-training.

\section{Conclusion}
In conclusion, we propose a novel task-aware framework, TA-GAT, focused on enhancing network generalization on both known and unseen task-based fMRI data. We will develop the architecture in a more general-purpose setting in the future, including extending the experiments to other datasets and exploring adding task awareness to unsupervised representation learning.

\section{Compliance with Ethical Standards}
This paper uses open-access fMRI data from WU-Minn HCP 1200 Subjects Data Release~\cite{hcp} and follows the HCP open access data use terms. Ethical approval was not required as confirmed by the license attached with the open access data.

\section{Acknowledgement}
This paper is supported under NIH grant R01NS035193.

\bibliographystyle{IEEEbib}
\bibliography{paper}

\end{document}